\documentclass[aps,prl,twocolumn,amsmath,amssymb,nofootinbib,superscriptaddress,floatfix,reprint,longbibliography]{revtex4-1}

\usepackage[dvips]{graphicx}
\usepackage{latexsym}
\usepackage{amsmath}
\usepackage{amsfonts}
\usepackage{amssymb}
\usepackage{bm}
\usepackage{color}
\usepackage{txfonts}
\usepackage{float}
\usepackage{url}
\usepackage[version=4]{mhchem}
\usepackage{braket}
\usepackage{booktabs}
\usepackage{siunitx}
\usepackage{mhchem}
\usepackage[colorlinks=true, urlcolor=blue, linkcolor=blue, citecolor=blue, pdftex]{hyperref}
\usepackage{ulem}
\begin{document}
\title{The structure of kagome superconductors \ce{CsV3Sb5} in the charge density wave states}
\author{Yuxin Wang}
\affiliation{Beijing National Laboratory for Condensed Matter Physics and Institute of Physics,
	Chinese Academy of Sciences, Beijing 100190, China}
\affiliation{School of Physical Sciences, University of Chinese Academy of Sciences, Beijing 100190, China}

\author{Tao Wu}
\affiliation{Hefei National Research Center for Physical Sciences at the Microscale, University of Science and
	Technology of China, Hefei, Anhui 230026, China}
\affiliation{CAS Key Laboratory of Strongly-coupled Quantum Matter Physics, Department of Physics, 
	University of Science and Technology of China, Hefei, Anhui 230026, China}
\author{Zheng Li}
\email{lizheng@iphy.ac.cn}
\affiliation{Beijing National Laboratory for Condensed Matter Physics and Institute of Physics,
	Chinese Academy of Sciences, Beijing 100190, China}
\affiliation{School of Physical Sciences, University of Chinese Academy of Sciences, Beijing 100190, China}

\author{Kun Jiang}
\email{jiangkun@iphy.ac.cn}
\affiliation{Beijing National Laboratory for Condensed Matter Physics and Institute of Physics,
	Chinese Academy of Sciences, Beijing 100190, China}
\affiliation{School of Physical Sciences, University of Chinese Academy of Sciences, Beijing 100190, China}

\author{Jiangping Hu}
\email{jphu@iphy.ac.cn}
\affiliation{Beijing National Laboratory for Condensed Matter Physics and Institute of Physics,
	Chinese Academy of Sciences, Beijing 100190, China}
\affiliation{Kavli Institute of Theoretical Sciences, University of Chinese Academy of Sciences,
	Beijing, 100190, China}

\date{\today}

\begin{abstract}
The structure of charge density wave states in AV$_3$Sb$_5$ (A = K, Rb, Cs) kagome superconductors remains elusive, with three possible $2a\times2a\times2c$ candidates: tri-hexagonal, star-of-David, and their mixture. In this study, we conducted a systematic first-principles investigation of the nuclear quadrupole resonance (NQR) and nuclear magnetic resonance (NMR) spectra for the $2a\times2a\times2c$ CsV$_3$Sb$_5$ structures. By comparing our simulations with experimental data, we have concluded that the NQR spectrum indicates the tri-hexagonal structure as the proper structure for CsV$_3$Sb$_5$ after its charge density wave phase transition. The NMR calculation results obtained from the tri-hexagonal structure are also consistent with the experimental data.
\end{abstract}
\maketitle

As one of the star materials in condensed matter physics, kagome materials are an important platform for investigating the interplay between correlation, topology, and geometric frustration. The recent discovery of kagome superconductors (SCs), AV$_3$Sb$_5$ (A=K,Rb,Cs), has taken the study of kagome physics to a new level \cite{ortiz19,ortiz20,ortiz21,lei,zywang,gaohj,kun_review,nematicity,musr1,topocdw,yuli,chenxh,AHE20,AHE22}. Many intriguing phenomena have emerged in AV$_3$Sb$_5$, including the anomalous Hall effect (AHE) \cite{chenxh,AHE20,AHE22}, pair density wave (PDW) \cite{gaohj}, electronic nematicity \cite{nematicity}, and possible evidence of time-reversal symmetry breaking from muon spin spectroscopy ($\mu$SR) measurements and optical rotation probes \cite{topocdw,musr1,yuli,feng,nanlin_wang,liangwu,balents,ronny,Nandkishore}.

In addition to the superconductor transition at $T_c$ around 1 $\sim$ 2 K, AV$_3$Sb$_5$ also exhibits another structure, a charge density wave (CDW) intertwined with a first-order transition at $T_s$ around 80 $\sim$ 100 K \cite{ortiz20,ortiz21,lei}. Various techniques have been employed to determine the low-temperature structure of AV$_3$Sb$_5$, including high-resolution X-ray diffraction (XRD) \cite{miaohu1,miaohu2,ortiz_prx,ortiz20,ortiz21,PhysRevMaterials.7.024806}, nuclear magnetic resonance (NMR) and nuclear quadrupole resonance (NQR) \cite{zhengli,nmr_tao,nqr,zhourui}, Raman spectroscopy \cite{miaohu1}, second harmonic generation (SHG) \cite{yuli}, scanning tunneling microscopy (STM) \cite{topocdw,zywang,Ilija}, and angle-resolved photoemission spectroscopy (ARPES) \cite{zhouxj}, etc. However, the true structure of AV$_3$Sb$_5$ remains controversial, and even similar experimental results can be interpreted in conflicting ways \cite{nqr,zhourui}. Therefore, theoretical interpretation and numerical simulations of the experiments are urgently needed.

In this study, we systematically investigated the nuclear quadrupole resonance (NQR) and nuclear magnetic resonance (NMR) spectra for various possible structures of AV$_3$Sb$_5$ using first-principles calculations. Our simulations generated distinct spectral shapes for a range of structures. By comparing these simulated spectra with experimental measurements, we have identified the tri-hexagonal (TrH) structure as the most likely content, rather than the star-of-David (SoD) structure and the recently proposed SoD-TrH mixture structure.

\begin{figure}[!htbp]
\centering
\includegraphics[width=8.6cm]{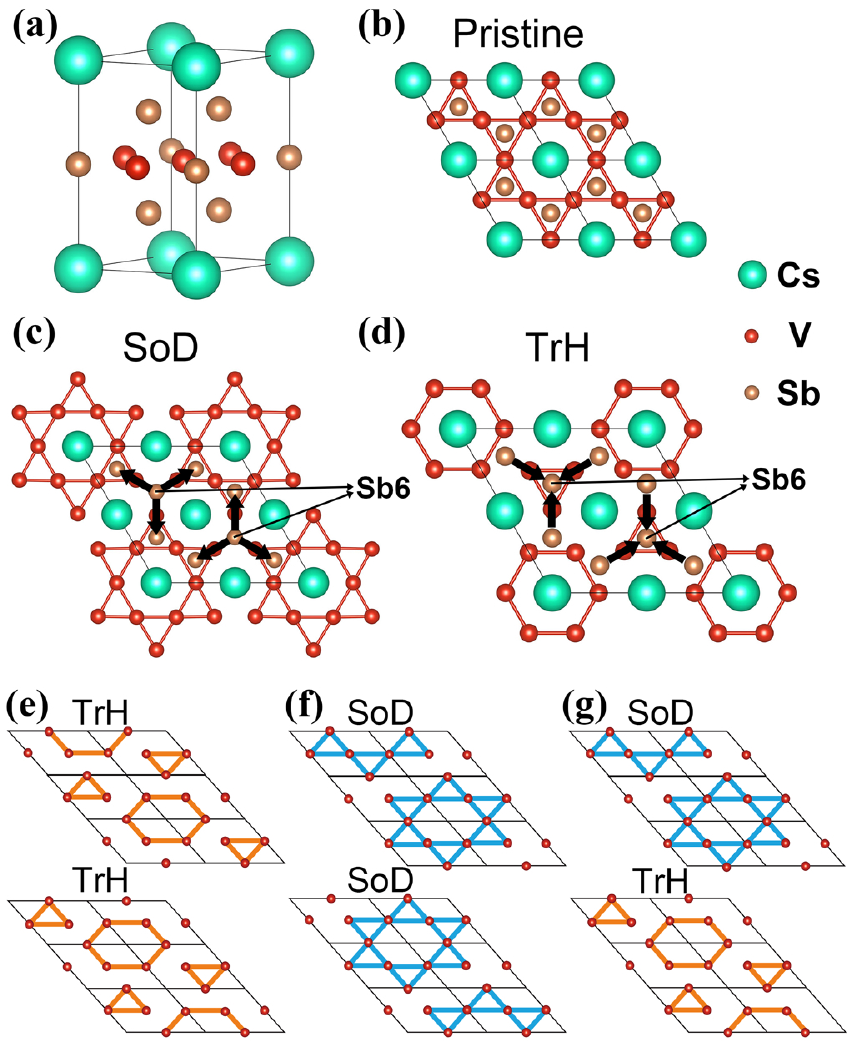}
\caption{(a) Axonometric view of the pristine crystal structure of \ce{CsV3Sb5}. (b) Top view of the pristine crystal structure.(c) Top view of the $2a\times2a$ SoD structure. (d) Top view of the $2a\times2a$ TrH structure. The moving directions of the three V atoms directly below Sb6 are also shown in (c),(d). (e) $2a\times2a\times2c$ pure TrH. (f) $2a\times2a\times2c$ pure SoD. (g) $2a\times2a\times2c$ TrH/SoD mix structure. Only the V kagome layer is shown in figures (e)-(g).}\label{fig;CDW pattern}
\end{figure}

The pristine AV$_3$Sb$_5$ structure is a layered structure of V-Sb sheets intercalated by A atoms with space group $P6/mmm$ \cite{ortiz19,ortiz20}, as shown in Fig. \ref{fig;CDW pattern}(a,b). In the V-Sb layer, the V atoms form a standard kagome lattice with additional Sb atoms located at the hexagonal center of the V kagome lattice. Above and below the V-Sb layer, there are two honeycomb lattice planes formed by out-of-plane Sb atoms situated above and below the centers of the V triangles in the kagome plane. The remaining A atoms form another triangular lattice in addition to the above layers.

Using ab initio calculations, two negative energy soft modes were identified around the M and L points in the phonon spectrum \cite{binghai,brian_anderson1,Andrzej_prb}. Based on the vibration pattern of these soft modes, two possible distorted structures for the V layers were proposed: the tri-hexagonal (TrH) and star-of-David (SoD) patterns \cite{binghai,brian_anderson1,Andrzej_prb}. In the SoD structure, V atoms are shrunk into a star-of-David or hexagram pattern, as shown in Fig. \ref{fig;CDW pattern}(c). The TrH structure, on the other hand, consists of two different components: small triangles and large hexagons, as illustrated in Fig. \ref{fig;CDW pattern}(d). The small triangles form a honeycomb lattice, while the large hexagons form a triangular lattice. Both distortions enlarge the unit cell to a $2a\times2a$ unit cell, as experimentally confirmed \cite{topocdw,miaohu1,miaohu2,ortiz_prx,PhysRevMaterials.7.024806}. From a point group perspective, both structures still have the D$_{6h}$ point group. Therefore, experimental techniques based on symmetry principles cannot directly differentiate between them. However, NQR and NMR, which probe the local environment of the atoms, provides a unique way to determine the structure of AV$_3$Sb$_5$.

In addition to the in-plane distortion, AV$_3$Sb$_5$ also undergoes a transition along the $c$-direction. The structural transitions in \ce{KV3Sb5} and \ce{RbV3Sb5} have been experimentally confirmed to be $2a\times2a\times2c$ \cite{topocdw,miaohu1}. However, the structure of \ce{CsV3Sb5} is still under debate, with proposed structures of $2a\times2a\times2c$ or $2a\times2a\times4c$ \cite{miaohu1,miaohu2,ortiz_prx,PhysRevMaterials.7.024806}. Since our calculations cannot settle this debate, we will focus only on the $2a\times2a\times2c$ structure in our discussion. Within this scope, we will use Cs as an example. We also performed NQR calculations for K and Rb (Tab. \ref{tab;exp K TrH NQR}-\ref{tab;exp Rb SoD NQR}), and find that the different alkali metal atoms does not impact our main results. With various proposals, we can construct three $2a\times2a\times2c$ structures: TrH/TrH, SoD/SoD, and TrH/SoD mixture, as illustrated in Fig. \ref{fig;CDW pattern}(e)-(g). 
In all the cases, we shift the $C_6$ center by one lattice constant to fulfill the $2c$ modulation between each layer. 
In the following discussion, we will only label them as TrH, SoD and Mix to simplify our notations.

In NQR and NMR spectra, the number, height, and position of peaks are key pieces of information for determining the crystal structure. The number of peaks reflects how many kinds of non-equivalent atoms are present in the material. The ratio of the heights of the peaks should be exactly equal to the ratio of the number of each kind of atom. The positions of the peaks reflect the local structural information of each kind of atom in the material. Therefore, a systematic analysis of the crystal structure is first needed. 

Due to the distortion in the $c$-direction, the six-fold rotation symmetry is broken. The pure TrH and pure SoD bilayer structures belong to space group $Fmmm$, while the Mix structure belongs to space group $Cmmm$. In the pristine structure above $T_s$, the three \ce{V} atoms are equivalent, while the five \ce{Sb} atoms are divided into two groups: one in-plane \ce{Sb}1 atom and four out-of-plane \ce{Sb}2 atoms.

\begin{figure}[!htbp]
\centering
\includegraphics[width=8.6cm]{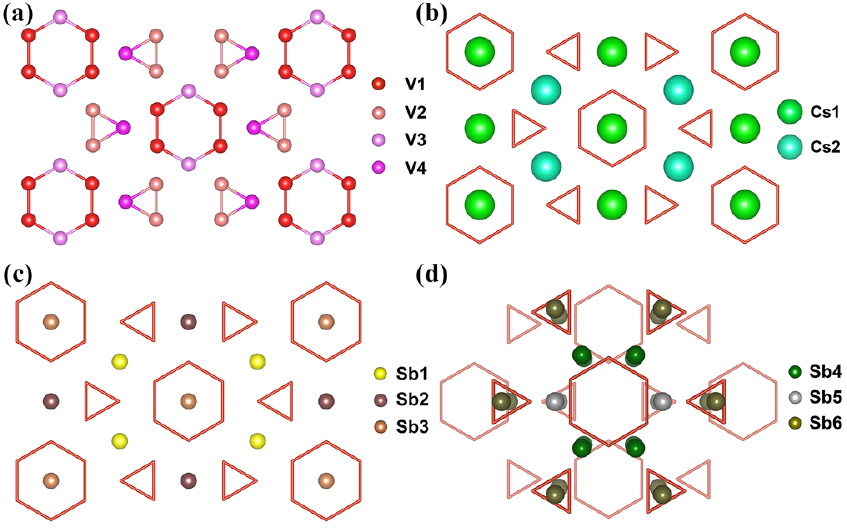}
\caption{Non-equivalent atoms of TrH. (a) Non-equivalent V atoms. (b) Non-equivalent Cs atoms. (c),(d) Non-equivalent Sb atoms. In order to make the figures clear, in figure (a), only one V kagome plane is shown, in figure (b), only one V kagome plane and the closest Cs plane are shown, in figure (c), only one V kagome plane and Sb1-3 atoms in the plane are shown, in figure (d), only two V kagome planes and Sb4-6 atoms adjacent to the upper V kagome plane are shown.}
\label{fig;TrH atoms}
\end{figure}

After the charge density wave phase transition, the equivalent class of atoms in AV$_3$Sb$_5$ largely changes. In the $2a\times2a\times2c$ TrH structure, there are two types of Cs atoms, four types of V atoms, and six types of Sb atoms (as shown in Fig. \ref{fig;TrH atoms} and Fig. \ref{fig;Non-equivalent atoms}(a)). The ratio of Cs atoms is Cs1:Cs2=1:1, while the ratio of V atoms is V1:V2:V3:V4=2:2:1:1, and the ratio of Sb atoms becomes Sb1:Sb2:Sb3:Sb4:Sb5:Sb6=2:1:1:8:4:4. V1 and V3 belong to the large hexagon, while V2 and V4 belong to the triangle in the TrH configuration. Sb1, Sb2, and Sb3 are located in the kagome plane and derive from Sb1 in the pristine structure, while Sb4, Sb5, and Sb6 are out of the plane and derive from Sb2 in the pristine structure. Sb4 and Sb5 are next to the hexagon, and Sb5 is directly above the V-triangle. Sb6 is also directly above the triangle, but it is much closer than Sb5.

In the $2a\times2a\times2c$ SoD structure, the situation is the same as TrH (as shown in Fig. \ref{fig;Non-equivalent atoms}(b)). Despite the size of the distortion, the only difference between the two structures is that the displacement direction of all V atoms in SoD is opposite to that in TrH \cite{binghai}. For the $2a\times2a\times2c$ Mix structure, the pattern of adjacent kagome planes is different. Hence, there are more different kinds of non-equivalent atoms than TrH and SoD. For Cs atoms, the number becomes three. For V and Sb, these numbers are doubled (as shown in Fig. \ref{fig;Non-equivalent atoms}(c)). The ratio of Cs atoms becomes Cs1:Cs2:Cs3=1:2:1. For V atoms, this ratio is V1:V2:V3:V4:V5:V6:V7:V8=2:2:1:1:2:2:1:1, and for Sb atoms, the ratio is Sb1:Sb2:Sb3:Sb4:Sb5:
Sb6:Sb7:Sb8:Sb9:Sb10:Sb11:Sb12=2:1:1:8:4:4:2:1:1:8:4:4.

Equipped with the above information, we can now present our computational results of the NQR spectra and compare them with the experimental results \cite{nqr,zhourui}. The elementary principle of NQR is based on the electric quadrupole moment $Q$ of the nucleus. This quadrupole moment arises from the non-spherical charge distribution of a nucleus with a spin quantum number $I$ greater than 1/2 \cite{Suits2006}. The quadrupole moment couples to the electric field gradient generated by the electronic bonds in the surrounding environment, resulting in the splitting of the nucleus energies. These energy splittings can be detected through radio-frequency (RF) magnetic fields, as in NMR spectroscopy \cite{Suits2006}. Therefore, NQR can be utilized to determine the distortion in the crystal structure.

The quadrupole interaction is described by an effective Hamiltonian \cite{slichter1996principles}:
\begin{align}
H_{Q}=\frac{eQV_{zz}}{4I(2I-1)}[(3I_{z}^{2}-I^{2})+\frac{1}{2}\eta(I_{+}^{2}+I_{-}^{2})],
\end{align}
where $V_{zz}$ is defined as the maximum eigenvalue of the electric field gradient tensor, and $\eta=\lvert({V_{xx}-V_{yy}})/{V_{zz}}\rvert$ is the asymmetry parameter. We focus on the experimentally available \ce{^{121}Sb} in the NQR spectra. For \ce{^{121}Sb}, $Q=\SI{-54.3d-30}{m^2}$ and $I=5/2$. Since $I=5/2$, two peaks can be detected experimentally for each kind of non-equivalent \ce{Sb} atom, belonging to $\ket{\pm1/2}\leftrightarrow\ket{\pm3/2}$ and $\ket{\pm3/2}\leftrightarrow\ket{\pm5/2}$ transitions. Here, we only consider the transition $\ket{\pm1/2}\leftrightarrow\ket{\pm3/2}$.

Another important quantity in NQR is the quadrupole resonance frequency $\nu_Q$, which is defined as $\nu_Q = \frac{3eQV_{zz}}{2I(2I-1)h}$ \cite{slichter1996principles}. The energy difference between the two states is denoted as $\Delta E$, and $f$ satisfies $hf = \Delta E$ corresponding to the frequency. When $\eta = 0$, the absolute value of the quadrupole resonance frequency is equal to the frequency corresponding to the transition $\ket{\pm1/2}\leftrightarrow\ket{\pm3/2}$. However, in most cases, $\eta$ is a small value, and second-order perturbation theory can be used to calculate $f$ when $\eta \neq 0$ (see Appendix):
\begin{align}
f = \left(1 + \frac{59}{54}\eta^2\right)|\nu_Q|.
\end{align}

For the first-principles calculations, we utilize the all-electron augmented plane wave (APW) basis set \cite{martin2020electronic,singh2006planewaves,sjostedt2000alternative} implemented in the WIEN2k package \cite{blaha2020wien2k}. The NQR calculation in WIEN2k is described in Chapter 6.4 of reference \cite{schwarz2010electronic}.

\begin{figure}[!htbp]
\centering
\includegraphics[width=8.6cm]{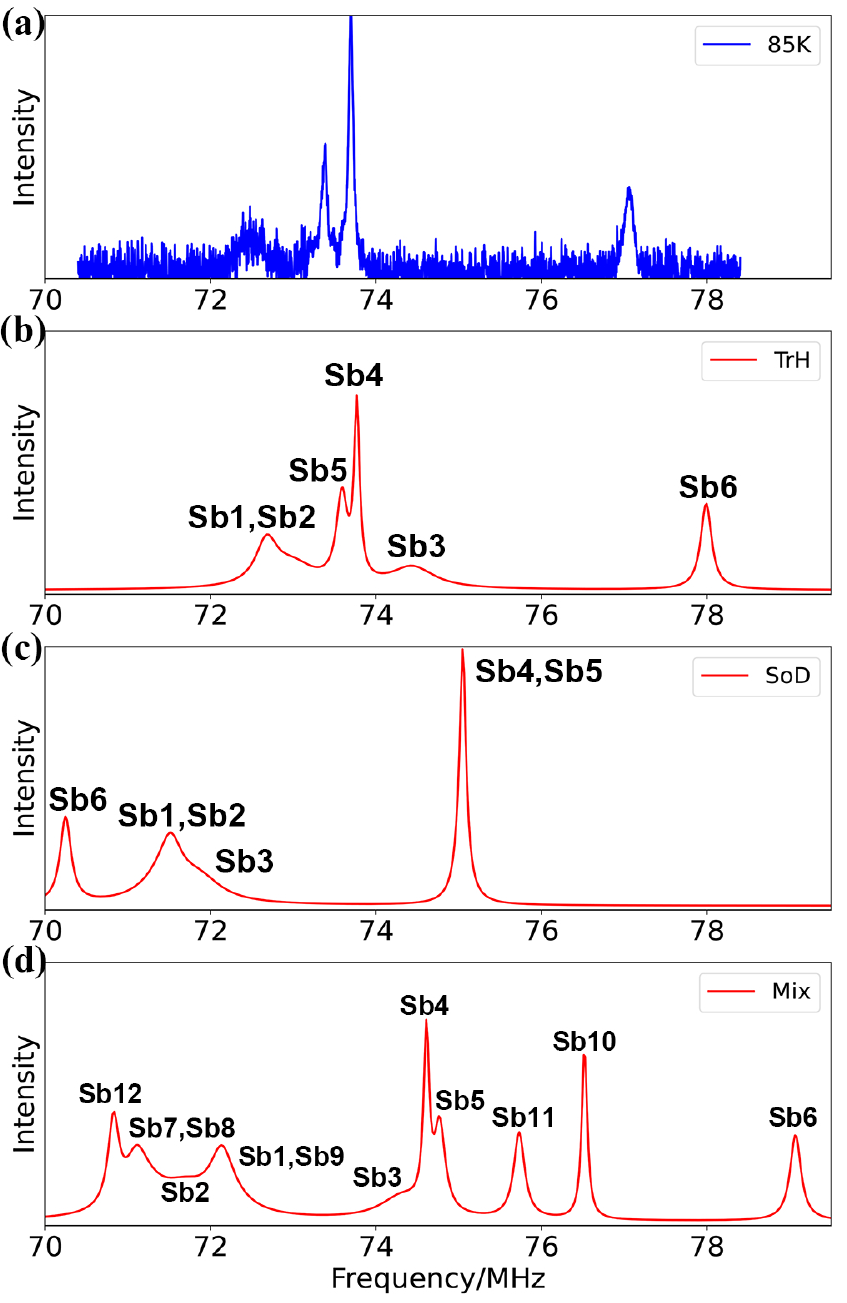}
\caption{The NQR spectra for experimental and calculated results. (a) Experimental spectrum at 85K (below $T_{s}$) \cite{nqr}. (b) Calculated spectrum of relaxed $2a\times2a\times2c$ TrH. (c) Calculated spectrum of relaxed $2a\times2a\times2c$ SoD. (d) Calculated spectrum of relaxed $2a\times2a\times2c$ mixture structure. For comparison purposes, the calculated results are shifted by +4.7 MHz.}\label{fig;NQR}
\end{figure}

We found that the calculated NQR spectra for the four different crystal structures in Fig. \ref{fig;CDW pattern}(a),(e)-(g) depend on the lattice parameters and the the lengths of each chemical bond in each structure. However, we need to point out that the main feature used to distinguish SoD from TrH does not depend on these factors, which will be discussed later. Therefore, we first need to find suitable crystal structures. For the undistorted pristine structure, we used the experimental structure. For the three candidate distorted structures, since there are no completely uncontroversial experimental data available, one option is to use the optimized results from density functional theory (DFT) calculations. The DFT structure optimization test showed that the lattice parameters in the kagome plane hardly changed (less than $0.01\AA$ compared to twice the undistorted experimental values), and the lattice parameter in the direction perpendicular to the plane did not change much (less than $0.05\AA$ compared to twice the undistorted experimental values). Therefore, as a reliable approximation, we took the lattice parameters of these three $2a\times2a\times2c$ structures to be twice the undistorted experimental values, and then optimized the internal coordinates. In the main text, we used the relaxed crystal structures to calculate the NQR spectra, which are shown in Fig. \ref{fig;NQR}(b)-(d). We then compared them with the experimental results \cite{nqr} (See Fig. \ref{fig;NQR}(a)). Recently, another study also provided some experimental refined structure models using single-crystal X-ray crystallographic refinements \cite{PhysRevMaterials.7.024806}. These experimental structures were also used to calculate the NQR parameters, which are shown in Tab. \ref{tab;exp TrH NQR}-\ref{tab;exp SoD NQR}. The results for the pristine structure are shown in Fig. \ref{fig;0.01-0.01 Fig NQR}(a,b). All the detailed data, including $V_{zz}$, $\nu_{Q}$, and $\eta$, can be found in the Appendix. In Fig. \ref{fig;NQR} and Fig. \ref{fig;0.01-0.01 Fig NQR}, the calculated data were broadened using the following Lorentzian function:
\begin{align}
y=\frac{2}{\pi}\frac{w}{4(x-x_{c})^{2}+w^{2}},
\end{align}
where $w=\frac{2}{\pi y_{c}}$ and ($x_{c}$,$y_{c}$) is the coordinate of the top of the Lorentzian peak. Here, $x_{c}$ is the frequency and $y_{c}$ is the intensity, which is proportional to the number of atoms.

As shown in Fig. \ref{fig;NQR}, the NQR spectrum of the $2a\times2a\times2c$ TrH structure exhibits a closer match to the experimental results, albeit with a constant frequency shift. However, it's important to note that the decision to exclude the other two structures was not solely based on the resemblance between the calculated and experimental spectra. Here, we will first examine the main distinguishing feature in the calculated NQR spectra of TrH and SoD, which is insensitive to lattice parameters and specific atomic positions. Based on this feature, we will then determine which type of NQR spectrum the experimental results belong to.

The main distinguishing feature between the calculated NQR spectra of TrH and SoD is the relative positions of the Sb4, Sb5, and Sb6 peaks, all of which arise from Sb2 in the pristine structure. These three kinds of Sb atoms are located outside the kagome plane, with each Sb atom having three nearest neighboring V atoms inside the adjacent kagome plane. In TrH, the Sb6 peak appears at a higher frequency than Sb4 and Sb5 (Fig. \ref{fig;NQR}(b)). On the other hand, for SoD, the situation is reversed (Fig. \ref{fig;NQR}(c)). This difference can be explained by the displacement directions of the three V atoms closest to Sb6. In TrH, the three V atoms are closer to Sb6 and form a triangle (Fig. \ref{fig;CDW pattern}(d) and Fig. \ref{fig;TrH atoms}(d)), whereas in SoD, the three V atoms are further away from Sb6 (Fig. \ref{fig;CDW pattern}(c)). However, for any Sb4 or Sb5 atom, two of the three V atoms move in a direction perpendicular to the V-Sb bond. Therefore, in the TrH structure, Sb6 experiences a larger electric field gradient than Sb4 and Sb5, resulting in a higher frequency of the Sb6 peak than the Sb4 and Sb5 peaks. In contrast, in the SoD structure, Sb6 experiences a smaller electric field gradient than Sb4 and Sb5, resulting in a lower frequency of the Sb6 peak than the Sb4 and Sb5 peaks. This result is caused by the inherent characteristics of the two structures and does not depend on lattice parameters or bond lengths. For both optimized DFT structures , experimental structures and other structures, this result can be observed (See Appendix). To further confirm the effect of the displacements of the three closest V atoms on the position of the Sb6 peak in the NQR spectra, we fixed the displacement of other V atoms and only varied the displacements of V atoms closest to Sb6. As expect, our results show that when the displacement of V atoms is greater (closer to Sb6), the Sb6 peak shifts to a higher frequency (Tab. \ref{tab;Sb6 NQR}).

After analyzing the computational results, we need to determine which case the experimental NQR spectrum corresponds to. This requires us to be able to assign the peaks in the experimental results to the corresponding Sb atoms, especially Sb4, Sb5, and Sb6. Here, we will use the more convincing indicators of peak height and asymmetry parameters ($\eta$) to accomplish this task. The calculated $\eta$ for both TrH and SoD are shown in Tab. \ref{tab;cal Asymmetry parameters}, while the experimental results are listed in Tab. \ref{tab;exp Asymmetry parameters}. Based on the number of atoms and the $\eta$ values, we determine that the highest peak and its adjacent peak (peak 3 and peak 2 in Tab. \ref{tab;exp Asymmetry parameters}) in Fig. \ref{fig;NQR}(a) should originate from Sb4 and Sb5. In previous experimental reports, the peak with the highest frequency ($\sim$ 77 MHz in Fig. \ref{fig;NQR}(a), peak 4 in Tab. \ref{tab;exp Asymmetry parameters}) has been considered to originate from Sb1-Sb3 which all arise from Sb1 in pristine structure due to its close-to-zero $\eta$ \cite{nqr,zhourui}. Here, we argue that this peak should not come from Sb1-Sb3. From the calculation results of DFT relaxed structures, only Sb3 and Sb6 satisfy the condition of $\eta$ equals to zero. However, the proportion of Sb3 is only 1/8 of Sb4, so it cannot produce such a high peak.
Regarding the experimental structure reported in Ref.\cite{PhysRevMaterials.7.024806}, the computational results indicate that only the Sb6 peak has a $\eta$ value close to zero (see Tab. \ref{tab;cal Asymmetry parameters} or Tab. \ref{tab;exp TrH NQR}-\ref{tab;exp SoD NQR}). In conclusion, this peak is more likely to originate from Sb6. Consequently, since the frequency of the Sb6 peak is higher than that of the Sb4 and Sb5 peaks in the experimental spectrum, we believe that TrH is a more plausible structure than SoD.

Finally, the calculated NQR spectrum of the Mix structure is simply a superposition of those of TrH and SoD (see Fig. \ref{fig;NQR}(d)). Thus, based solely on the number of peaks, Mix does not seem to be a suitable structure.

\begin{table}[!htbp] 
\centering
\caption{Asymmetry parameters $\eta$ for Sb atoms in calculated NQR speatra.}
\begin{ruledtabular}
\begin{tabular}{ccccccc}
&Sb1&Sb2&Sb3&Sb4&Sb5&Sb6 \\
\midrule
Cal.(pristine)&0&0&-&-&-&- \\
Cal.(relaxed TrH)&0.069&0.068&0&0.129&0.131&0.003 \\
Cal.(relaxed SoD)&0.045&0.047&0&0.093&0.09&0 \\
Cal.(exp. TrH\cite{PhysRevMaterials.7.024806})&0.035&0.022&0.037&0.04&0.039&0.007 \\
Cal.(exp. SoD\cite{PhysRevMaterials.7.024806})&0.036&0.022&0.038&0.035&0.041&0.001 
\end{tabular}
\end{ruledtabular}
\label{tab;cal Asymmetry parameters}
\end{table}

\begin{table}[!htbp] 
\centering
\caption{Asymmetry parameters $\eta$ for Sb atoms in experimental NQR speatra. The frequency of the peak increases from left to right.}
\begin{ruledtabular}
\begin{tabular}{ccccc}
&Peak1&Peak2&Peak3&Peak4 \\
\midrule
Exp.(95K)\cite{nqr}&0&0&-&- \\
Exp.(85K)\cite{nqr}&0.04&0.07&0.07&0 \\

\end{tabular}
\end{ruledtabular}
\label{tab;exp Asymmetry parameters}
\end{table}

To further support the conclusion that the CDW pattern below $T_s$ corresponds to TrH, we calculated the NMR knight shifts of Cs and V and compared them with experimental results \cite{zhengli,nmr_tao}.

When a uniform external magnetic field $\textbf{B}_{\textbf{ext}}$ is applied, the nuclear magnetic moments interact with it, which results in the nuclear Zeeman effect. The electron magnetic moments also interact with the nuclear magnetic moments, inducing an effective magnetic field $\textbf{B}_{\textbf{ind}}(\textbf{R})$ at position $\textbf{R}$. When $\textbf{B}_{\textbf{ext}}$ is not very strong, the relationship between the two fields is linear, given by:
\begin{align}
\textbf{B}_{\textbf{ind}}(\textbf{R})=K(\textbf{R})\textbf{B}_{\textbf{ext}},
\end{align}
where $K$ is known as the NMR knight shift tensor and can be divided into the orbital and spin parts \cite{slichter1996principles}, corresponding to the electron orbital magnetic moment and the electron spin magnetic moment, respectively. In experimental measurements, the external magnetic field is applied perpendicular to the kagome plane, along the $z$ direction. Hence, the experimental knight shift data corresponds to the $K_{zz}$ component. The experimental and calculated NMR knight shift values $K_{zz}$ are presented in Tab. \ref{tab;NMR}.

Compared to NQR, the NMR knight shift is more sensitive to the crystal structure. Our tests show that even a slight change in the V-V bond length can lead to a large change in the spin part of the knight shift, resulting in significant deviation from experimental results. Therefore, a relaxed crystal structure that still has a certain gap with the real structure is not suitable for NMR calculation. We either use the experimental $2a\times2a\times2c$ TrH structure in ref \cite{PhysRevMaterials.7.024806} or manually adjust the distortion. The results of the experimental structure are shown in Tab. \ref{tab;exp NMR}. We find that the calculated NMR knight shift values are very close to the experimental values \cite{nmr_tao,zhengli}, but four different peaks appear, which is inconsistent with the experimental result that only has two peaks. Hence, we also try to use manually tuned structures in NMR calculations. Unlike the experimental structure, we keep the hexagons as regular hexagons and the triangles as regular triangles, although the space group symmetry of the crystal does not guarantee this. For example, in Tab. \ref{tab;NMR}, we list the results for 0.01-0.01 TrH. The first 0.01 represents the displacement of the V atoms that make up the hexagons, and the second 0.01 represents the displacement of the V atoms that make up the triangles, and the unit is angstrom. The NMR results of some other manually tuned structures are shown in Tab. \ref{tab;other NMR}. For the DFT relaxed structure, the distortion is around 0.045-0.08.

In order to maintain consistency of the structures used for NMR and NQR calculations, the manually tuned structure 0.01-0.01 TrH is also used to calculate the NQR spectra (See Tab. \ref{tab;detailed 0.01-0.01 TrH NQR} and Fig. \ref{fig;0.01-0.01 Fig NQR}(c,d)). As expected, this structure gives the same result like the DFT relaxed TrH structure.

\begin{table}[!htbp] 
\centering
\caption{NMR knight shift $K_{zz}$ for Cs and V atoms in pristine and manually adjusted 0.01-0.01 TrH structures. The first 0.01 represents that the displacement of the V atoms that make up the hexagon is 0.01\AA, and the second 0.01 represents the displacement of the V atoms that make up the triangle is also 0.01\AA.}
\begin{ruledtabular}
\begin{tabular}{ccccccc}
&Cs1&Cs2&V1&V2&V3&V4 \\
\midrule
Exp.(95K)\cite{nmr_tao,zhengli}/ppm&3300&-&4000&-&-&- \\
Cal.(Pristine)/ppm&2745&-&3856&-&-&- \\
Exp.(93K)\cite{nmr_tao,zhengli}/ppm&3400&2250&3700&4200&-&- \\
Cal.(0.01-0.01TrH)/ppm&3107&2514&3811&4030&3823&4071 
\end{tabular}
\end{ruledtabular}
\label{tab;NMR}
\end{table}

One important point to note is that the $2a\times2a\times2c$ TrH crystal structure contains four types of non-equivalent V atoms, yet only two NMR peaks were experimentally observed below $T_{s}$. The NMR calculation results of the experimental structure reported in ref \cite{PhysRevMaterials.7.024806} do not lead to this conclusion. However, from the calculated results of manually tuned structures, the four V atoms can be divided into two groups: one group consists of V1 and V3, which belong to the hexagon, while the other group consists of V2 and V4, which belong to the triangle. The knight shifts for each group are so close that they cannot be distinguished in experiments, resulting in only two distinct NMR peaks being observed.

From Tab. \ref{tab;NMR} and Tab. \ref{tab;detailed NMR}, we can see that the calculated NMR knight shift values using manually tuned structures are very close to the experimental NMR results \cite{nmr_tao,zhengli}. The pristine structure used in the calculation is the experimental structure, and the calculated knight shift values are also close to the experimental values. Therefore, we have reason to believe that the real CDW structure has V atom displacements ranging from about 0.01$\AA$ to 0.015$\AA$ in both hexagons and triangles, or V-V bond lengths of about 2.735$\AA$ in hexagons and 2.725$\AA$ in triangles. Notably, the experimental $2a\times2a\times2c$ TrH structure in ref \cite{PhysRevMaterials.7.024806} has V-V bond lengths very close to those we have proposed here.

For both pristine and representative 0.01-0.01 TrH structures, the total NMR Knight shift is decomposed into the orbital term, spin Fermi contact term, and spin dipolar term. The detailed values are presented in Tab. \ref{tab;detailed NMR}. From Tab. \ref{tab;detailed NMR}, it is evident that the main contribution to the NMR knight shift for Cs atoms is the spin Fermi contact term, which may be due to the large paramagnetic contribution from the spin-polarized valence s electrons of Cs. However, for V atoms, the main contribution comes from the orbital term, which is also consistent with the experimental findings \cite{nmr_tao}. It is worth noting that the spin contact term is diamagnetic for V atoms, which is uncommon except for transition metal elements. This is due to the external magnetic field inducing a sizable V 3d spin-magnetic moment, which introduces a large core polarization of opposite sign \cite{laskowski2015nmr,ebert1986real}. Because of the anisotropy of the crystal field for V atoms, the spin dipolar term also has a significant contribution mainly from V 3d electrons. Overall, our NMR calculations also support the TrH as the correct CDW structure below $T_{s}$.

\begin{table}[!htbp] 
\centering
\caption{Decomposition NMR knight shift into orbital term($K_{o}$), spin Fermi contact term($K_{fc}$), and spin dipolar($K_{sd}$) term for pristine and 0.01-0.01 TrH structures.}
\begin{ruledtabular}
\begin{tabular}{ccccc}
&$K_{o}$&$K_{fc}$&$K_{sd}$&Total \\
\midrule
Pristine-Cs1/ppm&-5318&8118&-55&2745 \\
Pristine-V1/ppm&5613&-2366&609&3856 \\
TrH-Cs1/ppm&-5276&8395&-12&3107 \\
TrH-Cs2/ppm&-5293&7859&-52&2514 \\
TrH-V1/ppm&5644&-2390&557&3811 \\
TrH-V2/ppm&5743&-2245&532&4030 \\
TrH-V3/ppm&5635&-2406&593&3822 \\
TrH-V4/ppm&5753&-2288&606&4071
\end{tabular}
\end{ruledtabular}
\label{tab;detailed NMR}
\end{table}

In summary, we conducted a systematic first-principles study of \ce{CsV3Sb5} using NQR and NMR spectroscopy. We considered three possible $2a\times2a\times2c$ structures, TrH, SoD, Mix, and the pristine structure to obtain $^{121}$Sb NQR spectra by calculating the NQR electric field gradients and asymmetry parameters. The main feature in $^{121}$Sb NQR spectra is determined by the relative positions of the Sb4 and Sb5 peaks to the Sb6 peak, which all come from Sb2 in the pristine structure. Comparing our results with experimental measurements \cite{nqr,zhourui}, we conclude that the TrH structure is more likely to be the correct structure.
To further confirm this result, we calculated the NMR knight shift for the TrH structure and the pristine structure. Although the TrH structure has four non-equivalent V atoms, if the hexagons and triangles mentioned are regular hexagons and equilateral triangles, respectively, they can be divided into two groups based on whether they belong to the hexagon or triangle. Within each group, two kinds of V atoms have a very close knight shift, which can account for only two V NMR peaks appearing in the experiments below $T_{s}$ \cite{nmr_tao,zhengli}. Therefore, our calculations demonstrate that within the $2a\times2a\times2c$ range, TrH should be the right structure below $T_{s}$ for \ce{CsV3Sb5}, and possibly for all \ce{AV3Sb5} materials, but not the SoD \cite{zhourui} or the coexistence of TrH and SoD \cite{ortiz_prx,wushangfei}.
We also want to mention that our calculations do not settle the debate between $2a\times2a\times2c$ and $2a\times2a\times4c$ in \ce{CsV3Sb5} \cite{miaohu2,ortiz_prx,PhysRevMaterials.7.024806,pengyy}, which requires further investigation.

Note that, when finalizing this work, we became aware of another work working on \ce{RbV3Sb5} with similar results \cite{rbv3sb_nqr}.

This work is supported by the National Natural Science Foundation of China (Grant No. NSFC-11888101, No. NSFC-12174428),  the Strategic Priority Research Program of Chinese Academy of Sciences (Grant No. XDB28000000), and the Chinese Academy of Sciences through the Youth Innovation Promotion Association (Grant No. 2022YSBR-048).

\bibliography{NQR_NMR.bbl}


\clearpage
\appendix
\begin{center}
\textbf{\large Appendix}
\end{center}
\renewcommand{\thetable}{S\arabic{table}}
\setcounter{table}{0}
\renewcommand{\thefigure}{S\arabic{figure}}
\setcounter{figure}{0}

\section{THEORETICAL APPROACH}
\subsection{NQR}
Here, we will outline the basic theory for calculating the relevant parameters of nuclear quadrupole resonance (NQR), such as the electric field gradient $V_{zz}$ and the asymmetry parameter $\eta$ \cite{slichter1996principles}. When we do not consider the nucleus as a point charge, the nucleus in different states will have different interaction energies with the external potential field under the same external potential field. Classically, the interaction energy $E$ of a charged nucleus with a charge density $\rho(\textbf{r})$ with an external potential $V(\textbf{r})$ can be described as:

\begin{align}
E=\int\rho(\textbf{r})V(\textbf{r})d^3r\label{NQR interaction}.
\end{align}

We expand $V(\textbf{r})$ to second order in a Taylor's series about the origin:
\begin{align}
V(\textbf{r})=V(\textbf{0})+\textbf{r}\cdot\nabla V\mid_{\textbf{r}=\textbf{0}}+\frac{1}{2}\sum_{i,j=1}^{3}x_{i}x_{j}\frac{\partial^{2}V}{\partial x_{i}\partial x_{j}}\mid_{\textbf{r}=\textbf{0}}\label{Taylor series of V}.
\end{align}

We set the position of a nucleus as the origin. When we put equation \ref{Taylor series of V} into equation \ref{NQR interaction},the first order term $\int\rho(\textbf{r})\textbf{r}\cdot\nabla V\mid_{\textbf{r}=\textbf{0}}d^3r$ vanishes if the nuclear states possess the definite parity. Fortunately, all experimental evidence supports this contention.

It is convenient to define another quantities $Q_{ij}$:
\begin{align}
Q_{ij}=\int(3x_{i}x_{j}-\delta_{ij}r^2)\rho(\textbf{r})d^3r.
\end{align}
The external potential must satisfy Laplace equation:$\nabla^{2}V=0$. Defining $\frac{\partial^{2}V}{\partial x_{i}\partial x_{j}}\mid_{\textbf{r}=\textbf{0}}=V_{ij}$, the second order term,which is also called the quadurpole energy $E^{(2)}$, is given by:
\begin{align}
E^{(2)}=\frac{1}{6}\sum_{i,j=1}^{3}V_{ij}Q_{ij}.
\end{align}

It can be quantized by replacing the charge density $\rho(\textbf{r})$ with the corresponding quantum mechanical operator:
\begin{align}
\rho(\textbf{r})=e\sum_{k}\delta(\textbf{r}-\textbf{r}_{\textbf{k}}).
\end{align}
Summing over k here means summing over all protons, because we no longer think of nuclei as point charges.  In the subspace formed by the total angular momentum eigenstates of the nucleus, using Wigner-Eckart theorem, the quadrupole energy can be described by an effective hamiltonian:
\begin{align}
H_{Q}=\frac{eQV_{zz}}{4I(2I-1)}[(3I_{z}^{2}-I^{2})+\frac{1}{2}\eta(I_{+}^{2}+I_{-}^{2})].
\end{align} 
The matrix $V_{ij}$ is symmetric so that it can be simplified by choosing a set of principal axes in which the matrix is diagonal. $V_{zz}$ is defined as the maximum eigenvalue of the matrix $V_{ij}$. $\eta=\lvert({V_{xx}-V_{yy}})/{V_{zz}}\rvert$ is asymmetry parameter, in which $V_{xx}$ and $V_{yy}$ are another two eigenvalues. These quantities can be easily computed using DFT. $I$ is the quantum number of nuclear total spin angular momentum and $Q$ is the nuclear quadrupole moment, which can be determined by experiments.  

For \ce{^{121}Sb}, $I=5/2$. Here, we condider the energy difference corresponding to the transition $\ket{1/2}\leftrightarrow\ket{3/2}$ when $\eta\neq0$. $\ket{-1/2}\leftrightarrow\ket{-3/2}$ and $\ket{1/2}\leftrightarrow\ket{3/2}$ are the same.

When $\eta$ is small, $H^{'}=eQV_{zz}\eta(I_{+}^{2}+I_{-}^{2})/[8I(2I-1)]=h\nu_{Q}\eta(I_{+}^{2}+I_{-}^{2})/12$ can be regarded as a perturbation. The second order perturbation theory must be used because the first order perturbation of energy is zero:
\begin{align}
\begin{split}
E_{1/2}^{(2)}=&\frac{1}{144}h^{2}\nu_{Q}^{2}\eta^{2}\left(\frac{\lvert\bra{5/2}I_{+}^{2}\ket{1/2}\rvert^{2}}{E_{1/2}^{(0)}-E_{5/2}^{(0)}}+\frac{\lvert\bra{-3/2}I_{-}^{2}\ket{1/2}\rvert^{2}}{E_{1/2}^{(0)}-E_{-3/2}^{(0)}}\right)\\
=&-\frac{16}{27}h\nu_{Q}\eta^{2}
\end{split}
\end{align}
\begin{align}
E_{3/2}^{(2)}=&\frac{1}{144}h^{2}\nu_{Q}^{2}\eta^{2}\frac{\lvert\bra{-1/2}I_{-}^{2}\ket{3/2}\rvert^{2}}{E_{3/2}^{(0)}-E_{-1/2}^{(0)}}=\frac{1}{2}h\nu_{Q}\eta^{2}
\end{align}
So, the energy difference is:
\begin{align}
\Delta E=\lvert E_{1/2}^{(0)}+E_{1/2}^{(2)}-E_{3/2}^{(0)}-E_{3/2}^{(2)}\rvert=(1+\frac{59}{54}\eta^{2})h\lvert\nu_{Q}\rvert
\end{align}

\subsection{NMR}
As we describe in the main text, if we apply an uniform external magnetic field $\textbf{B}_{\textbf{ext}}$, it will cause an effective induced magnetic field $\textbf{B}_{\textbf{ind}}(\textbf{R})$ at the position \textbf{R}. If the $\textbf{B}_{\textbf{ext}}$ is not very strong, the relationship between the two is linear:
\begin{align}
\textbf{B}_{\textbf{ind}}(\textbf{R})=K(\textbf{R})\textbf{B}_{\textbf{ext}},
\end{align}
where $K$ is called NMR knight shift tensor and can be devided into orbital part and spin part. In WIEN2k, we usually using $\sigma=-K$ as NMR knight shift tensor.
\subsubsection{Orbital Part} 
In the framework of first order perturbation theory, the orbital part of the induced magnetic field $\textbf{B}_{\textbf{ind}}^{\textbf{orb}}$ can be calculated applying the Biot-Savart law \cite{slichter1996principles,laskowski2012calculations}:
\begin{align}
\textbf{B}_{\textbf{ind}}(\textbf{R})=\frac{1}{c}\int\textbf{j}_{\textbf{ind}}(\textbf{r})\times\frac{\textbf{R}-\textbf{r}}{\lvert\textbf{R}-\textbf{r}\rvert^{3}}d^{3}r.
\end{align}

The orbital motion of the electrons is affected by the $\textbf{B}_{\textbf{ext}}$. In the DFT framework, we need to solve the Kohn-Sham (KS) equation in the external magnetic field to acquire induced current density $\textbf{j}_{\textbf{ind}}(\textbf{r})$. The KS Hamiltonian in the magnetic field is:
\begin{align}
H=\frac{1}{2}[\textbf{p}+\frac{1}{c}\textbf{A}(\textbf{r})]^{2}+V(\textbf{r}),
\end{align}
where $\nabla\times\textbf{A}(\textbf{r})=\textbf{B}_{\textbf{ext}}$. In the symmetric gauge, $\textbf{A}(\textbf{r})=\frac{1}{2}\textbf{B}_{\textbf{ext}}\times\textbf{r}$. So the $\textbf{j}_{\textbf{ind}}(\textbf{r})$ is:
\begin{align}
\begin{split}
\textbf{j}_{\textbf{ind}}(\textbf{r})=&\sum_{o}[-\frac{1}{2}(\bra{\psi_{o}}\textbf{p}\ket{\textbf{r}}\braket{\textbf{r}|\psi_{o}}-\braket{\psi_{o}|\textbf{r}}\bra{\textbf{r}}\textbf{p}\ket{\psi_{o}})\\
&-\frac{\textbf{B}_{\textbf{ext}}\times\textbf{r}}{2c}\lvert\braket{\textbf{r}|\psi_{o}}\rvert^2]\label{induced current},
\end{split}
\end{align}
where $\ket{\psi_{o}}$ is the occupied KS orbital. Compared to the effective single particle potential $V(\textbf{r})$, $\textbf{B}_{\textbf{ext}}$ is small. Therefore, the second order term of $\textbf{B}_{\textbf{ext}}$ in the KS Hamiltonian can be ignored and the KS Hamiltonian becomes:
\begin{align}
H=\frac{1}{2}\textbf{p}^2+V(\textbf{r})+\frac{1}{2c}(\textbf{r}\times\textbf{p})\cdot\textbf{B}_{\textbf{ext}}.
\end{align}
The first order term $H^{(1)}=\frac{1}{2c}(\textbf{r}\times\textbf{p})\cdot\textbf{B}_{\textbf{ext}}$ can be regarded as a perturbation. If $\ket{\psi_{o}^{(0)}}$ is unperturbed KS orbital, using standard density functional perturbation theory (DFPT) \cite{baroni2001phonons}, the first order eigenstate is:
\begin{align}
\ket{\psi_{o}^{(1)}}=\sum_{e}\frac{\ket{\psi_{e}^{(0)}}\bra{\psi_{e}^{(0)}}\frac{1}{2c}(\textbf{r}\times\textbf{p})\cdot\textbf{B}_{\textbf{ext}}\ket{\psi_{o}^{(0)}}}{E_{o}-E_{e}}\label{first order eigenstate}.
\end{align}
The sum is running over all the empty KS orbitals. $E_{o}$ and $E_{e}$ are the eigenvalues of $\ket{\psi_{o}^{(0)}}$ and $\ket{\psi_{e}^{(0)}}$ with respect to the unpertubative Hamiltonian, respectively. Put the equation \ref{first order eigenstate} into \ref{induced current} and accurate to the first order of $\textbf{B}_{\textbf{ext}}$, the $\textbf{j}_{\textbf{ind}}(\textbf{r})$ is:
\begin{align}
\begin{split}
\textbf{j}_{\textbf{ind}}(\textbf{r})=&\sum_{o}[-\frac{1}{2}(\bra{\psi_{o}^{(1)}}\textbf{p}\ket{\textbf{r}}\braket{\textbf{r}|\psi_{o}^{(0)}}-\braket{\psi_{o}^{(0)}|\textbf{r}}\bra{\textbf{r}}\textbf{p}\ket{\psi_{o}^{(1)}})\\
&-\frac{\textbf{B}_{\textbf{ext}}\times\textbf{r}}{2c}\lvert\braket{\textbf{r}|\psi_{o}^{(0)}}\rvert^2].
\end{split}
\end{align}
\subsubsection{Spin Part}
The spin part of the induced magnetic field $\textbf{B}_{\textbf{ind}}^{\textbf{spin}}$ is described as \cite{novak2006calculation,slichter1996principles,laskowski2015nmr,blugel1987hyperfine}:
\begin{align}
\textbf{B}_{\textbf{ind}}^{\textbf{spin}}=\frac{8\pi}{3}\textbf{m}_{\textbf{av}}+\int\frac{S(r)}{r^{3}}\frac{3(\textbf{m}(\textbf{r})\cdot\textbf{r})\textbf{r}-\textbf{m}(\textbf{r})r^{2}}{r^{2}}d^{3}r,
\end{align}
where the first term on the right-hand side of the equation is the spin Fermi contact term, while the second term is the spin dipolar term. These terms can be derived by simplifying the Dirac equation in the presence of an electromagnetic field. The spin Fermi contact term is related to the average spin magnetization density ($\textbf{m}_{\textbf{av}}$) over a region near the nucleus with a diameter equal to the Thomson radius. This term contributes only when the wavefunction is non-zero at the nucleus, which is the case only for wavefunctions corresponding to the s electrons in the material. Therefore, the spin Fermi contact term arises primarily from the s electrons and dominates the spin part of the calculation. In contrast, electrons other than s electrons can contribute to the spin dipolar term, which vanishes for high symmetry structures. The relativistic correction factor $S(r)$ and the spin magnetization density $\textbf{m}(\textbf{r})$ also appear in the equation.

In WIEN2k, only the contribution from within the atomic sphere is considered because the contribution from the interstitial region is negligible. If we only consider the contribution from within atomic sphere for spin dipolar term, it can be simplified as \cite{novak2006calculation,laskowski2017computational,abragam2012electron}:
\begin{align}
\begin{split}
\textbf{B}_{\textbf{ind}}^{\textbf{sd}}=&\frac{4\mu_{B}}{(2l+3)(2l-1)}\sum_{o}\bra{\psi_{o}}\frac{S(r)}{r^{3}}[l(l+1)\textbf{s}\\&-\frac{3}{2}(\textbf{l}\cdot\textbf{s})\textbf{l}-\frac{3}{2}\textbf{l}(\textbf{l}\cdot\textbf{s})]\ket{\psi_{o}},
\end{split}
\end{align}
where the summation over all occupied states.
\section{COMPUTATIONAL METHOD}
In this study, we employed the all-electron augmented plane wave (APW) basis set \cite{martin2020electronic,singh2006planewaves,sjostedt2000alternative} as implemented in the WIEN2k package \cite{blaha2020wien2k}. Within the APW method, the unit cell is partitioned into non-overlapping atomic spheres and an interstitial region. The wave function is expressed as a linear combination of spherical harmonics inside the spheres and as a linear combination of plane waves in the interstitial region. The energy cutoff, $RK_{max}$, was set to the default value of 7, which ensures convergence in both NQR and NMR calculations.

In the NQR calculation, we used 18191 k-points and 4104 k-points in the first Brillouin zone for the pristine and three candidate $2a\times2a\times2c$ structures, respectively. 

In the NMR calculation, we computed the orbital and spin contributions separately. For the orbital part, we employed a gauge-invariant perturbation method \cite{laskowski2012calculations}, while for the spin part, we used a direct self-consistent all-electron approach \cite{novak2006calculation}. A brief description of these methods is provided above. The value of the "smearing" parameter has an impact on the final converged results \cite{laskowski2015nmr}, so we established a benchmark using the pristine structure, for which we have experimental data. We chose a "smearing" parameter of 2 mRy, which yielded results consistent with the experiments. To ensure both speed and accuracy, we used 97556 and 6540 k-points in the first Brillouin zone for the pristine and TrH structures, respectively. In the NMR spin part calculation, we used 96715 and 8646 k-points. We implemented spin-orbit coupling (SOC) in spin part calculation to set the quantization axis perpendicular to the kagome plane, in agreement with the experiments.

\section{NQR/NMR RESULTS SUPPLEMENT}
\begin{figure*}[!htbp]
\centering
\includegraphics[width=17.2cm]{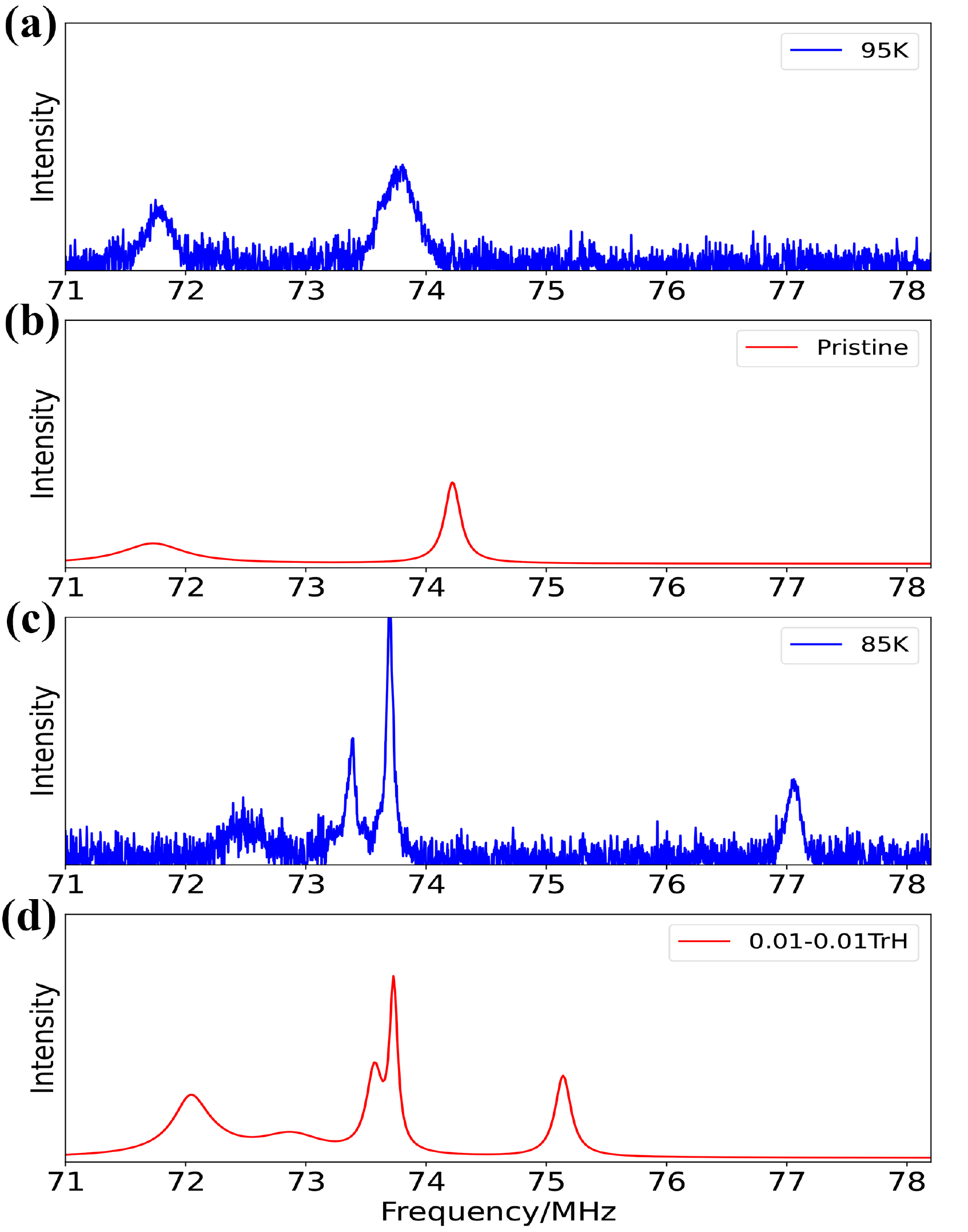}
\caption{The NQR spectra for experimental results and manually adjusted 0.01-0.01 TrH. (a) The experimental spectrum \cite{nqr} at 95K (above $T_{s}$). (b) The calculated spectrum of the pristine structure. (c) The experimental spectrum \cite{nqr} at 85K (below $T_{s}$). (d) The calculated spectrum of the 0.01-0.01 TrH structure. For comparison purposes, the calculated results are shifted by +5.75MHz.}\label{fig;0.01-0.01 Fig NQR}
\end{figure*}

We present the detailed NQR parameters for the pristine and three relaxed candidate CDW structures in Tab. \ref{tab;detailed pristine NQR}-\ref{tab;detailed Mix NQR}. The calculated NQR parameters for the experimental $2a\times2a\times2c$ TrH and SoD structures are shown in Tab. \ref{tab;exp TrH NQR} and \ref{tab;exp SoD NQR}, respectively. In Tab. \ref{tab;Sb6 NQR}, we present the NQR results for Sb6 in different manually adjusted structures, which support our statement in the main text. Additionally, we provide the NQR results for the manually adjusted 0.01-0.01 TrH structure in Tab. \ref{tab;detailed 0.01-0.01 TrH NQR} and compare the experimental results at 95K with the calculated results of the pristine structure and the experimental results at 85K with the calculated results of the 0.01-0.01 TrH in Fig. \ref{fig;0.01-0.01 Fig NQR}. To facilitate comparison, the calculated results are shifted by +5.75MHz. 

The NQR results of \ce{KV3Sb5} and \ce{RbV3Sb5} are also calculated using the experimental structures from ref \cite{PhysRevMaterials.7.024806}, which can be seen in Tab. \ref{tab;exp K TrH NQR}-\ref{tab;exp Rb SoD NQR}.

Furthermore, we report the NMR Knight shift values for the experimental $2a\times2a\times2c$ TrH in Tab. \ref{tab;exp NMR} and for other manually tuned TrH structures in Tab. \ref{tab;other NMR}.

\section{NON-EQUIVALENT ATOMS}
\begin{figure*}[!htbp]
\centering
\includegraphics[width=17.2cm]{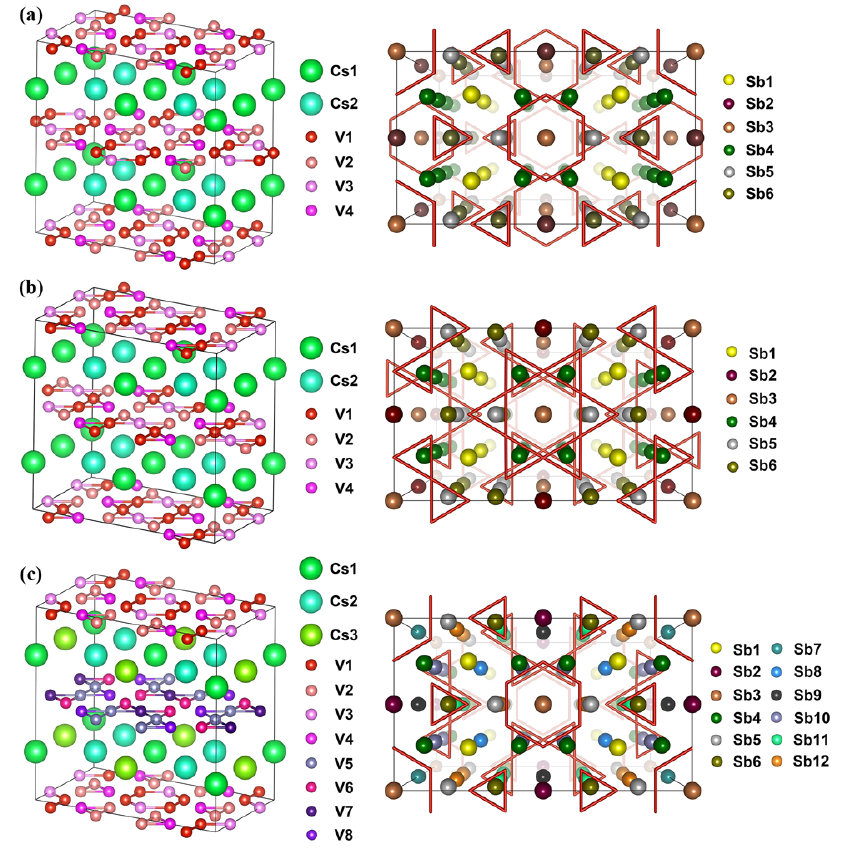}
\caption{The different kinds of non-equivalent atoms of \ce{CsV3Sb5}.(a)$2a\times2a\times2c$ pure TrH.(b)$2a\times2a\times2c$ pure SoD.(c)$2a\times2a\times2c$ mixed structure.The left half of the graph contains only Cs atoms and V atoms, and the right half of the graph contains only Sb atoms and bonds between V atoms.}\label{fig;Non-equivalent atoms}
\end{figure*}

Fig. \ref{fig;Non-equivalent atoms} displays the different non-equivalent atoms in the three candidate CDW structures, indicated by different colors.

\begin{table}[!htbp] 
\centering
\caption{Detailed NQR parameters for pristine structure.}
\begin{ruledtabular}
\begin{tabular}{cccc}
&$\lvert V_{zz}\rvert \times10^{21}\SI{}{V/m^{2}}$&$\eta$&$f$/MHz\\
\midrule
Sb1&33.547&0&65.979 \\
Sb2&34.813&0&68.47 
\end{tabular}
\end{ruledtabular}
\label{tab;detailed pristine NQR}
\end{table}

\begin{table}[!htbp] 
\centering
\caption{Detailed NQR parameters for relaxed $2a\times2a\times2c$ TrH.}
\begin{ruledtabular}
\begin{tabular}{cccc}
&$\lvert V_{zz}\rvert \times10^{21}\SI{}{V/m^{2}}$&$\eta$&$f$/MHz\\
\midrule
Sb1&34.385&0.069&67.977 \\
Sb2&34.549&0.068&68.297 \\
Sb3&35.461&0&69.743 \\
Sb4&34.475&0.13&69.068 \\
Sb5&34.398&0.129&68.891 \\
Sb6&37.261&0.003&73.285 
\end{tabular}
\end{ruledtabular}
\label{tab;detailed TrH NQR}
\end{table}

\begin{table}[!htbp] 
\centering
\caption{Detailed NQR parameters for relaxed $2a\times2a\times2c$ SoD.}
\begin{ruledtabular}
\begin{tabular}{cccc}
&$\lvert V_{zz}\rvert \times10^{21}\SI{}{V/m^{2}}$&$\eta$&$f$/MHz\\
\midrule
Sb1&33.9&0.045&66.825 \\
Sb2&33.844&0.047&66.727 \\
Sb3&34.151&0&67.167 \\
Sb4&35.431&0.093&70.347 \\
Sb5&35.454&0.09&70.345 \\
Sb6&33.327&0&65.547 
\end{tabular}
\end{ruledtabular}
\label{tab;detailed SoD NQR}
\end{table}

\begin{table}[!htbp] 
\centering
\caption{Detailed NQR parameters for relaxed $2a\times2a\times2c$ Mix.}
\begin{ruledtabular}
\begin{tabular}{cccc}
&$\lvert V_{zz}\rvert \times10^{21}\SI{}{V/m^{2}}$&$\eta$&$f$/MHz\\
\midrule
Sb1&34.041&0.082&67.439 \\
Sb2&33.816&0.081&66.982 \\
Sb3&35.393&0&69.611 \\
Sb4&34.893&0.131&69.915 \\
Sb5&34.962&0.132&70.069 \\
Sb6&37.814&0.002&74.372 \\
Sb7&33.659&0.054&66.41 \\
Sb8&33.762&0.044&66.544 \\
Sb9&34.295&0.006&67.454 \\
Sb10&36.145&0.097&71.817 \\
Sb11&35.665&0.107&71.027 \\
Sb12&33.605&0.021&66.127 
\end{tabular}
\end{ruledtabular}
\label{tab;detailed Mix NQR}
\end{table}

\begin{table}[!htbp] 
\centering
\caption{Detailed NQR parameters for experimental $2a\times2a\times2c$ TrH.}
\begin{ruledtabular}
\begin{tabular}{cccc}
&$\lvert V_{zz}\rvert \times10^{21}\SI{}{V/m^{2}}$&$\eta$&$f$/MHz\\
\midrule
Sb1&33.006&0.035&65.004 \\
Sb2&32.998&0.022&64.935 \\
Sb3&33.155&0.037&65.308 \\
Sb4&34.059&0.04&67.106 \\
Sb5&34.041&0.039&67.063 \\
Sb6&35.319&0.007&69.468 
\end{tabular}
\end{ruledtabular}
\label{tab;exp TrH NQR}
\end{table}

\begin{table}[!htbp] 
\centering
\caption{Detailed NQR parameters for experimental $2a\times2a\times2c$ SoD.}
\begin{ruledtabular}
\begin{tabular}{cccc}
&$\lvert V_{zz}\rvert \times10^{21}\SI{}{V/m^{2}}$&$\eta$&$f$/MHz\\
\midrule
Sb1&33.059&0.036&65.112 \\
Sb2&33.005&0.022&64.946 \\
Sb3&32.971&0.038&64.949 \\
Sb4&34.657&0.035&68.252 \\
Sb5&34.655&0.041&68.287 \\
Sb6&33.478&0.001&65.844 
\end{tabular}
\end{ruledtabular}
\label{tab;exp SoD NQR}
\end{table}

\begin{table}[!htbp] 
\centering
\caption{NQR parameters for Sb6 in different manually adjusted $2a\times2a\times2c$ TrH structures.}
\begin{ruledtabular}
\begin{tabular}{cccc}
&$\lvert V_{zz}\rvert \times10^{21}\SI{}{V/m^{2}}$&$\eta$&$f$/MHz\\
\midrule
0.01-0.01&35.127&0.005&69.09 \\
0.01-0.015&35.302&0.007&69.436 \\
0.01-0.02&35.52&0.007&69.862 \\
0.01-0.045&36.128&0.002&71.055 \\
0.01-0.06&36.297&0&71.387 
\end{tabular}
\end{ruledtabular}
\label{tab;Sb6 NQR}
\end{table}

\begin{table}[!htbp] 
\centering
\caption{Detailed NQR parameters for manually adjusted 0.01-0.01 $2a\times2a\times2c$ TrH.}
\begin{ruledtabular}
\begin{tabular}{cccc}
&$\lvert V_{zz}\rvert \times10^{21}\SI{}{V/m^{2}}$&$\eta$&$f$/MHz\\
\midrule
Sb1&33.538&0.023&66.285 \\
Sb2&33.57&0.023&66.348 \\
Sb3&33.987&0&67.134 \\
Sb4&34.367&0.035&67.975 \\
Sb5&34.298&0.031&67.819 \\
Sb6&35.127&0.005&69.387 
\end{tabular}
\end{ruledtabular}
\label{tab;detailed 0.01-0.01 TrH NQR}
\end{table}

\begin{table}[!htbp] 
\centering
\caption{Detailed NQR parameters for experimental \ce{KV3Sb5} $2a\times2a\times2c$ TrH.}
\begin{ruledtabular}
\begin{tabular}{cccc}
&$\lvert V_{zz}\rvert \times10^{21}\SI{}{V/m^{2}}$&$\eta$&$f$/MHz\\
\midrule
Sb1&33.771&0.046&66.573 \\
Sb2&33.705&0.018&66.313 \\
Sb3&34.283&0.038&67.535 \\
Sb4&33.639&0.071&66.521 \\
Sb5&33.324&0.062&65.813 \\
Sb6&34.986&0.019&68.838 
\end{tabular}
\end{ruledtabular}
\label{tab;exp K TrH NQR}
\end{table}

\begin{table}[!htbp] 
\centering
\caption{Detailed NQR parameters for experimental \ce{KV3Sb5} $2a\times2a\times2c$ SoD.}
\begin{ruledtabular}
\begin{tabular}{cccc}
&$\lvert V_{zz}\rvert \times10^{21}\SI{}{V/m^{2}}$&$\eta$&$f$/MHz\\
\midrule
Sb1&33.939&0.056&66.98 \\
Sb2&33.977&0.02&66.855 \\
Sb3&33.711&0.003&66.303 \\
Sb4&34.074&0.06&67.277 \\
Sb5&34.452&0.063&68.053 \\
Sb6&32.841&0.012&64.601 
\end{tabular}
\end{ruledtabular}
\label{tab;exp K SoD NQR}
\end{table}

\begin{table}[!htbp] 
\centering
\caption{Detailed NQR parameters for experimental \ce{RbV3Sb5} $2a\times2a\times2c$ TrH.}
\begin{ruledtabular}
\begin{tabular}{cccc}
&$\lvert V_{zz}\rvert \times10^{21}\SI{}{V/m^{2}}$&$\eta$&$f$/MHz\\
\midrule
Sb1&32.961&0.063&65.106 \\
Sb2&33.014&0.015&64.947 \\
Sb3&33.756&0.032&66.467 \\
Sb4&34.147&0.093&67.795 \\
Sb5&33.982&0.093&67.467 \\
Sb6&36.207&0.003&71.211 
\end{tabular}
\end{ruledtabular}
\label{tab;exp Rb TrH NQR}
\end{table}

\begin{table}[!htbp] 
\centering
\caption{Detailed NQR parameters for experimental \ce{RbV3Sb5} $2a\times2a\times2c$ SoD.}
\begin{ruledtabular}
\begin{tabular}{cccc}
&$\lvert V_{zz}\rvert \times10^{21}\SI{}{V/m^{2}}$&$\eta$&$f$/MHz\\
\midrule
Sb1&33.195&0.064&65.577 \\
Sb2&33.105&0.01&65.117 \\
Sb3&32.926&0.031&64.828 \\
Sb4&34.965&0.081&68.769 \\
Sb5&35.192&0.069&69.573 \\
Sb6&33.147&0.01&65.201 
\end{tabular}
\end{ruledtabular}
\label{tab;exp Rb SoD NQR}
\end{table}

\begin{table}[!htbp] 
\centering
\caption{NMR knight shift $K_{zz}$ for Cs and V atoms in experimental TrH structure.}
\begin{ruledtabular}
\begin{tabular}{ccccccc}
&Cs1&Cs2&V1&V2&V3&V4 \\
\midrule
Cal.(Exp.TrH)/ppm&3470&2140&3966&4362&3163&3632
\end{tabular}
\end{ruledtabular}
\label{tab;exp NMR}
\end{table}

\begin{table}[!htbp] 
\centering
\caption{NMR knight shift $K_{zz}$ for Cs and V atoms in some other manually adjusted structures.}
\begin{ruledtabular}
\begin{tabular}{ccccccc}
&Cs1&Cs2&V1&V2&V3&V4 \\
\midrule
Cal.(0.01-0.015TrH)/ppm&3007&2162&3840&4124&3686&4037 \\
Cal.(0.015-0.01TrH)/ppm&2984&2378&3818&4068&3788&4060 \\
Cal.(0.015-0.015TrH)/ppm&2825&1991&3820&4163&3696&4063
\end{tabular}
\end{ruledtabular}
\label{tab;other NMR}
\end{table}

\newpage

\end{document}